\documentclass[twocolumn,groupedaddress,floatfix,amsmath,amssymb,superscriptaddress]{revtex4-2}
\usepackage{amsmath}
\usepackage{amssymb}
\usepackage{braket}
\usepackage{graphicx}
\usepackage[table]{xcolor}
\usepackage{verbatim}
\usepackage{float}
\usepackage{color}
\usepackage{siunitx}
\usepackage{gensymb}
\usepackage{bm}
\usepackage[table]{xcolor}
\usepackage{tabularx}
\usepackage{multirow}
\usepackage{footnote}
\usepackage{bbm}
\usepackage{booktabs}
\definecolor{BLUE}{RGB}{0, 0, 255}
\usepackage{adjustbox}
\usepackage{orcidlink}
\usepackage{siunitx}
\usepackage{rotating}
\usepackage{array}

\newcolumntype{Y}{>{\centering\arraybackslash}X}
\newcolumntype{C}[1]{>{\centering\arraybackslash}p{#1}}

\usepackage{physics}

\begin{document}
\title{Density Functional Theory for 2D Transition-Metal Dichalcogenides with Extended Hubbard Parameters }
\begin{abstract}
Accurate modelling of the band-edge electronic structure of transition metal dichalcogenides (TMDs), such as MoS$_2$, is essential for understanding their optical, electronic and spintronic properties and enabling future applications. Here, we supplement the standard PBE exchange-correlation functional with on-metal ($U$) and metal--chalcogen ($V$) Hubbard corrections, which can control the orbital composition of band-edge states and consequently improve the quantitative accuracy of various properties. For MoS$_2$, MoSe$_2$, WS$_2$, and WSe$_2$, we optimise the $U$ and $V$ parameters by comparison with experimentally measured spin--orbit splittings at the conduction and valence band edges, and show that these optimal choices also improve agreement of the computed band gap, and ferroelectric potential drop in rhombohedral bilayers, with experimentally measured values. This provides a highly efficient, scalable framework for quantitatively accurate, high-throughput calculations of monolayer and multilayer TMDs, suitable for implementation in 2D-materials databases, as well as large-scale electronic and spintronic device modelling.
\end{abstract}

\author{Xue Li\,
\orcidlink{0009-0009-9994-4717}}
\affiliation{Department of Physics and Astronomy, University of Manchester, Oxford Road, Manchester, M13 9PL, United Kingdom}
\affiliation{National Graphene Institute, University of Manchester, Booth St.\ E., Manchester, M13 9PL, United Kingdom}
\author{Suad Alshammari\,
\orcidlink{0009-0008-4780-8855}}
\affiliation{Department of Physics and Astronomy, University of Manchester, Oxford Road, Manchester, M13 9PL, United Kingdom}
\affiliation{National Graphene Institute, University of Manchester, Booth St.\ E., Manchester, M13 9PL, United Kingdom}
\affiliation{Physics Department, Faculty of Science, Northern Border University, Arar, Saudi Arabia}
\author{Igor Rozhansky\,
\orcidlink{0000-0001-9391-9304}}
\affiliation{Department of Physics and Astronomy, University of Manchester, Oxford Road, Manchester, M13 9PL, United Kingdom}
\affiliation{National Graphene Institute, University of Manchester, Booth St.\ E., Manchester, M13 9PL, United Kingdom}
\author{James G. McHugh\,
\orcidlink{0000-0001-8509-4883}}
\email{james.mchugh@manchester.ac.uk}
\affiliation{Department of Physics and Astronomy, University of Manchester, Oxford Road, Manchester, M13 9PL, United Kingdom}
\affiliation{National Graphene Institute, University of Manchester, Booth St.\ E., Manchester, M13 9PL, United Kingdom}
\author{Vladimir  Fal'ko\,
\orcidlink{0000-0003-0828-0310}}
\email{vladimir.falko@manchester.ac.uk}
\affiliation{Department of Physics and Astronomy, University of Manchester, Oxford Road, Manchester, M13 9PL, United Kingdom}
\affiliation{National Graphene Institute, University of Manchester, Booth St.\ E., Manchester, M13 9PL, United Kingdom}

\maketitle

\section{Introduction}
Nowadays, the importance of transition-metal dichalcogenides (TMDs) for semiconductor technologies is broadly recognised, winning them a prominent place in the IEEE International Roadmap for Devices and Systems \cite{IRDS2024Metrology}. For example, their atomically thin geometry enables intricate control of electrostatic gating, allowing aggressively scaled and low-power transistors, while weak 2D dielectric screening and quantum confinement in tandem favour strong electron–hole interactions and tunable optoelectronic response. Realizing this potential, by identifying multilayer configurations suitable for novel device applications, demands highly accurate prediction of electronic properties. This includes properties such as the energy offset between different band-extrema, spin-orbit splitting of the conduction- and valence-band (CB, VB) edges, as well as CB, VB interlayer coupling, and the ferroelectric potential drop in stacked bilayers.

\begin{figure*}[ht]
\centering    
\includegraphics[width=2\columnwidth]{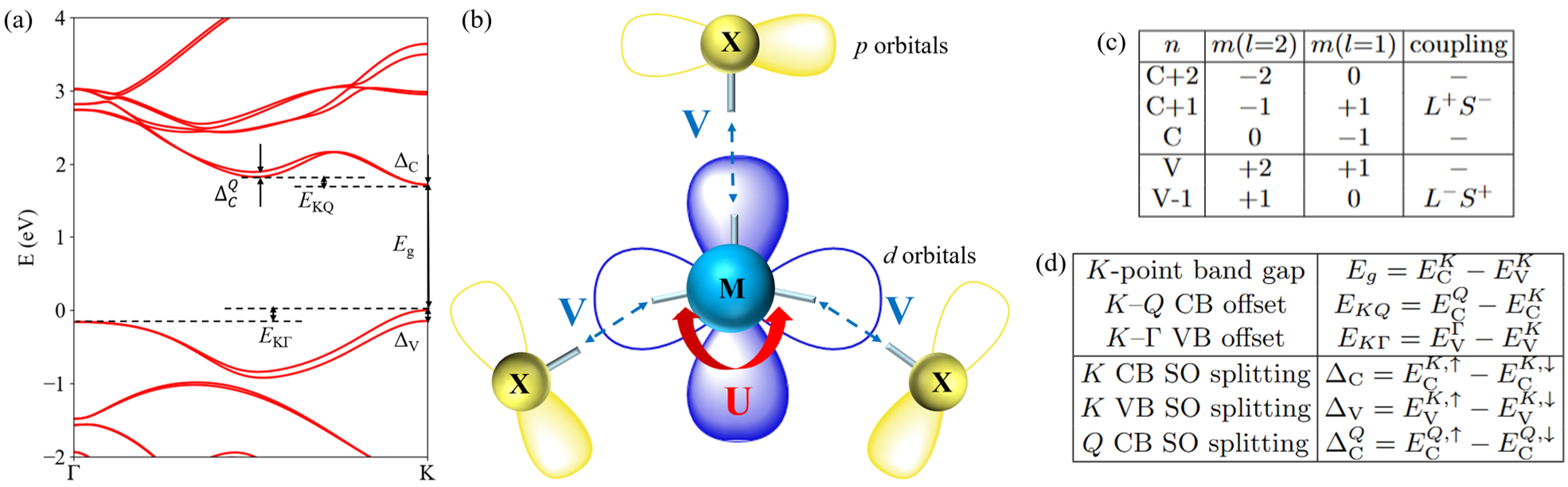}
\caption{(a) Overview of the band structure of monolayer TMDs (MX$_2$, exemplified for MoS$_2$) obtained from DFT calculations. Dispersion along $\Gamma-K$ line in the Brillouin zone, illustrating the various band offsets and SO splittings we consider in this work. (b) U is on-metal site Hubbard interaction, V is inter-site Hubbard interaction, and on-chalcogen Hubbard interaction is neglected. (c) Orbital composition of metal and chalcogen components of Bloch states, and allowed second-order contributions to SO splitting with the CB, $C$, at the K point in monolayer TMDs. (d) Key band-edge quantities used throughout this work.}
   \label{fig:band}
\end{figure*}

However, achieving quantitatively accurate predictions of these properties remains difficult for conventional density functional theory (DFT), as their precise values depend intimately on electronic states in which a delicate balance between self-interaction and delocalisation determines the mixing of transition-metal (M) and chalcogen (X) orbitals. The development of efficient and reliable first-principles methods capable of capturing this intricate band-edge physics across monolayer and multilayer systems will therefore be essential for guiding future materials design and accelerating the translation of TMDs from laboratory curiosities into practical technologies.

Apparently, standard DFT often under- or over-estimates orbital weights, leading to discrepancies between experimentally measured and DFT-computed parameters of TMDs. For example, in monolayer MoS$_2$, standard DFT calculations consistently yield a CB spin-orbit (SO) splitting $\Delta_C$ (Fig.~\ref{fig:band}(a)) of only a few meV ($\approx 3$ meV), almost regardless of the level of theory employed (PBE, G$_0$W$_0$) \cite{RodriguesPela2024}. This stands in clear contrast with experimental estimates of roughly $8.1$ meV \cite{Rozhansky2026}. Other properties that are sensitive to orbital weight show similar discrepancies. For example, the built-in interlayer electrostatic potential drop in rhombohedral bilayer MoS$_2$ is approximately $63$--$70$ mV according to DFT calculations \cite{McHugh2024, Li2025, Weston2022}, while experimental measurements suggest values in the range $50$--$58$ mV \cite{Weston2022, PhysRevX.12.041005, Yang2024}. More generally, standard DFT also struggles to reproduce the experimentally determined band gap. The direct optical and quasiparticle gaps of monolayer MoS$_2$ are approximately $1.8$ and $2.16$ eV, respectively, whereas PBE-GGA values consistently fall below both, at around $1.65$ eV \cite{RodriguesPela2024}, raising the question of whether this discrepancy can also be partially offset by changes in orbital composition.

Here, we employ the Hubbard DFT+$U$+$V$ approach \cite{korotin2026parameterizing,Haas2024,PhysRevB.103.045141} to improve the description of orbital composition. Following Campo and Cococcioni\cite{Campo2010}, the total energy is amended by both on-site Hubbard energy $E_U$ and inter-site Hubbard energy 
\begin{equation}
\begin{split}
E_{UV}={}&\frac{U}{2}\sum_{M,\sigma}\mathrm{Tr}\big[\mathbf{n}_{MM}^\sigma(1-\mathbf{n}_{MM}^\sigma)\big]\\
&-\frac{V}{2}\sum_{M,X,\sigma}\mathrm{Tr}\big[\mathbf{n}_{MX}^\sigma\mathbf{n}_{XM}^\sigma\big]
\end{split}
\end{equation}
where ${\bf n}_{MM}^\sigma$ is the on-site occupation matrix of localized orbitals on metal atom $M$ with spin $\sigma$, and ${\bf n}_{MX}^\sigma$ is the inter-site occupation matrix between metal atom $M$ and chalogen, $X$, orbitals, with spin $\sigma$ (see SI section C for details), with associated, on-site Hubbard interaction energies, $U$, and inter-site ($M$--$X$) $V$ interaction parameters. These two terms are illustrated schematically in Fig.~\ref{fig:band}(b).\cite{Campo2010,Mahajan2021,TancogneDejean2020,Yang2024,Haas2024}

Using this extended DFT approach, we compute band structures, as exemplified in Fig.~\ref{fig:band}(a) for monolayer MoS$_2$. The primary features of such electronic band structure, which we use to tune and benchmark the proposed approach, are illustrated in Fig.~\ref{fig:band}(a). This includes the band gap, $E_g$, conduction and valence band SO splittings at the $K$, $\Delta_{\rm C,V}$ and $Q$, $\Delta_{\rm C}^Q$, valleys, and the relative energy separations between the $K$, $Q$ and $\Gamma$ band extrema. Fig.~\ref{fig:band}(d) summarizes these band-edge quantities, and their corresponding definitions, which were used to extract them from raw band energies. These properties are compared to the available experimental data based on Rydberg
series analysis \cite{PhysRevLett.123.136801, Vaquero2020, PhysRevB.110.235413, PhysRevLett.113.076802, Hsu2019, Viner2021, Liu2021, Kapuciski2021}, that extends 
photoluminescence (PL) spectroscopy \cite{Park2018, Huang2015, Island2016, Tongay2012, PhysRevLett.105.136805, Chi2023, Robert2020, PhysRevLett.105.136805, Stier2016, Mak2012, Kozawa2014, Zhu2015, Zhang2023, Hanbicki2015, Ye2014, Zhao2012, Zeng2013, PhysRevB.107.245407, Mehew2017, Ugeda2014, PhysRevLett.114.037401, Grzeszczyk2020, Shepard2017, Tongay2012, Kim2016, Lu2019, Robert2020, Wu2021, Ross2013, PhysRevLett.113.026803, Zhao2012, Hanbicki2015, PhysRevLett.113.026803, Hanbicki2015, PhysRevB.107.245407, Lu2019}, and scanning tunneling spectroscopy\cite{PhysRevLett.113.066105, Zhang2014, Zhang2015, Chiu2015, PhysRevB.99.115434, Trainer2019, Huang2015, Pieli2021, Hill2016, Rigosi2016, Zeng2013, Thomas2024, Hsu2025, Ugeda2014, Wu2021} of those materials, as well as frequency analysis of Shubnikov–de Haas oscillations (SdHO) detecting the doping offset for the occupation of upper spin states for conduction band electrons \cite{Rozhansky2026, PhysRevLett.121.247701, PhysRevB.97.201407, Wang2016} and angle-resolved photoemission spectroscopy (ARPES) resolving spin states at the top of the valence band \cite{PhysRevLett.114.046802, Lee2021, Ehlen2018, PhysRevB.92.245442, Tanabe2016,Ulstrup2016, PhysRevB.97.155421, Zhang2013, Mo2016, Zhao2019, PhysRevB.101.165103, Zhang2016, Le2015}.

\section{Results}
\subsection{Hubbard modulation of orbital composition}
\begin{figure}[ht]
\centering    
\includegraphics[width=1\columnwidth]{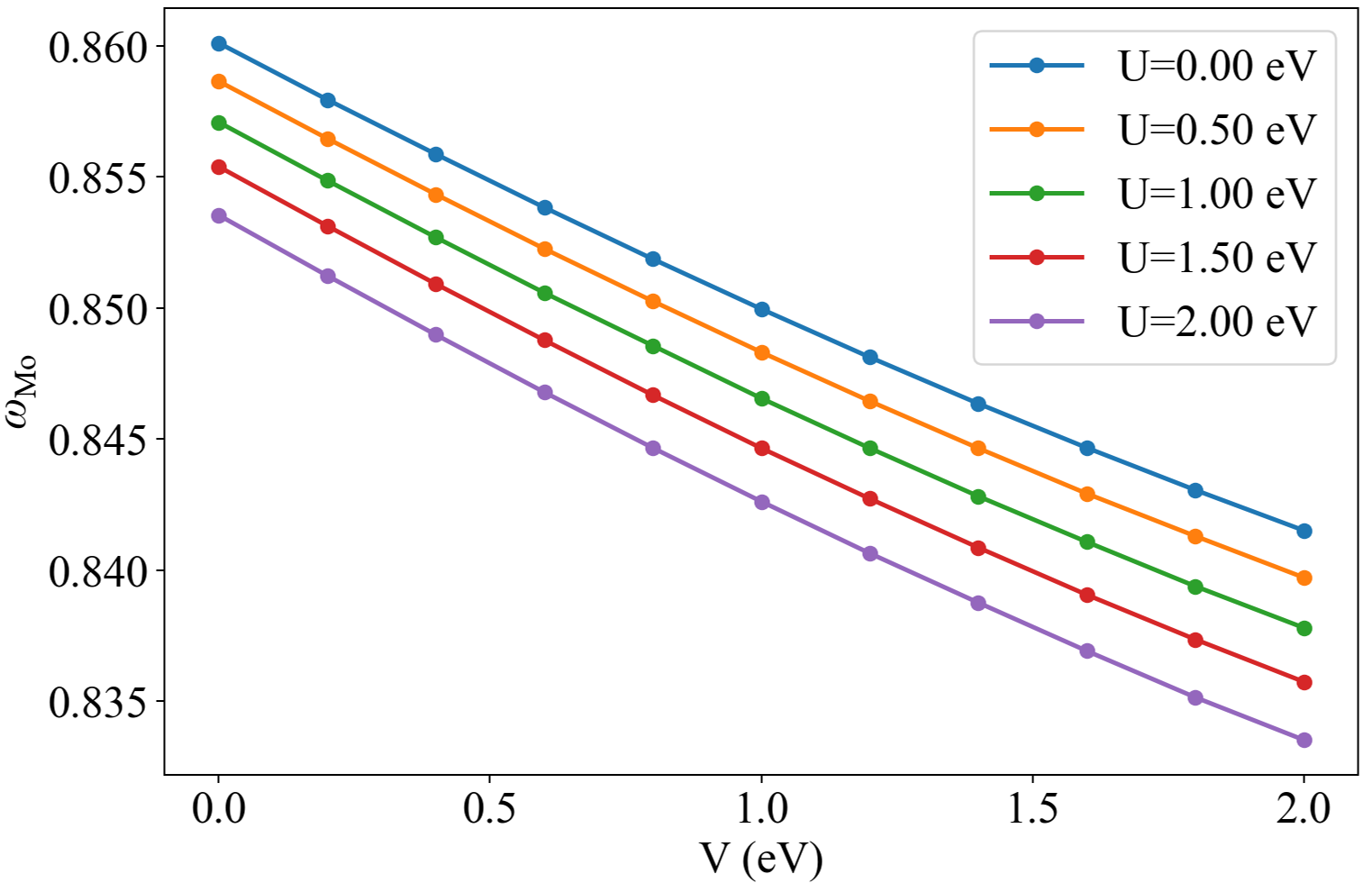}
    \caption{Dependence of the metal $w_{Mo}$ orbital weight of CB at $K$ point on the inter-site Hubbard parameter $V$ for different on-site Hubbard parameters $U$.}
    \label{fig:weight}
\end{figure}

Fig.~\ref{fig:weight} illustrates the dependence of the $K$-point CB-edge orbital weights on both Hubbard parameters. With increasing inter-site $V$, the Mo-$d$ weight, $w_{\rm Mo}$ decreases, with a corresponding increase in the S-$p$-orbital contribution, as $w_{\rm S} = 1-w_{\rm Mo}$. In addition, we note that on-site $U$ can also suppress $w_{\rm Mo}$, with a notably smaller effect than the inter-site term. This dependence can be understood within a minimal hybridisation picture, in which the $K$-point CB edge is composed of symmetry-compatible metal $d$ and chalcogen $p$ states. The orbital composition of this state is controlled jointly by the relative energetic alignment of each orbital manifold, and by inter-site $M$--$X$ hopping matrix elements. 

On-site Hubbard $U$ terms primarily contribute to the former, whereas the inter-site $V$ acts directly on the inter-site occupation matrix. The corresponding Hubbard potential contains off-diagonal terms of the form $-V_{MX}n^\sigma_{XM}$, which acts as a density-dependent correction to the effective inter-site $M$--$X$ coupling. This is consistent with the original DFT+$U$+$V$ formulation, in which the inter-site interaction counteracts the tendency of the on-site $U$ correction towards atomic localisation, stabilising hybridisation between neighbouring sites \cite{Campo2010,Mahajan2021}. For TMDs, we have confirmed this directly using Wannier90 interpolation of the DFT bands, which shows that the largest first-order effect of $V$ is to modify the nearest-neighbour $M$--$X$ hopping strength. It is the resulting redistribution of orbital weight which directly modifies the band-edge wave functions.

\subsection{Conduction \& valence band spin-orbit splitting}
Our initial point of comparison between theory and experiment is the SO splitting. To this end, we first review the physical mechanisms underlying the SO splitting of TMDs. To understand why such a correction can correct ths SO-splitting, we first review the physical processes governing the SO physics. In TMDs, the Zeeman SOC splits spin up and down states of band $n$ at a given momentum $k$. The total Hamiltonian is, 

\begin{align}
        &\hat{H}(\mathbf{k}) = \hat{H}_0(\mathbf{k}) +J_{X} \hat{\mathbf{L}}_X\cdot \hat{\mathbf{S}} + J_{M} \hat{\mathbf{L}}_M\cdot \hat{\mathbf{S}}, 
\end{align}
where $\hat{H}_0(\mathbf{k})$ is the ordinary scalar-relativistic Hamiltonian, and site-resolved SO operators of the form, $\hat{\mathbf{L}}\cdot \hat{\mathbf{S}} = \hat{{L}}^z \hat{{S}^z} + \frac{1}{2}(\hat{{L}}^+ \hat{{S}^-} + \hat{{L}}^- \hat{{S}^+}),$ act as a perturbation to the scalar-relativistic electronic structure, $\hat{\mathbf{L}}_{X/M}$ is the on-chalcogen/metal orbital angular momentum operator, $\hat{\mathbf{S}}$ is the spin operator, and $J_M, J_X$ are the SO constants corresponding to the transition metal and chalcogen atom, respectively. 

The first-order contribution to SO splitting in each band therefore arises from the on-site SO of the constituent atoms and the orbital weights of those atoms in a respective band. At the VBM, the electronic states are dominated by the metal $d_{x^2-y^2}$ and $d_{xy}$ orbitals ($m_z=2$), which, with the large metallic SO parameter gives a large, first-order splitting. In contrast, the CBM mainly derives from the metal $d_{z^2}$ orbital, which has $m_z=0$, and hence there is no metallic first-order contribution to the SO splitting. The delicate interplay between the first-order chalcogen, and second-order metal contributions is therefore important in determining the observed CB slitting.

The first and second order contribution to the SO splitting reads
\begin{gather}
\Delta_C \equiv E_C^\uparrow - E_C^\downarrow = \Delta_C^{(1)} + \Delta_C^{(2)},\nonumber\\
\Delta_C^{(1)} = w_M J_M\langle d_i|L^z_M S^z|d_i\rangle + w_X J_X\langle p_j|L^z_X S^z|p_j\rangle,\nonumber\\
\Delta_C^{(2)} \approx \frac{3}{2}\,\frac{J_M^2\,\omega_M^C\omega_M^{C+1}}{E_C-E_{C+1}},
\end{gather}
where $w_{\mathrm{M}}, w_{\mathrm{X}}$ are the weights of the metal $d$-orbital and chalcogen $p$-orbital components in the  wave function. The second-order terms in the CB, have contributions from the $C+3$, $C+1$ $V-1$, and $V-6$ bands, the closest of which are shown Fig.~\ref{fig:band}(c). These terms generally enter with different signs depending on angular momentum and energy offset, but the most important contribution comes from the nearest energy $C+1$ band, which has opposite sign to the first-order chalcogen contribution. See SI for further details.

\begin{figure*}[ht]
\centering    
\includegraphics[width=2\columnwidth]{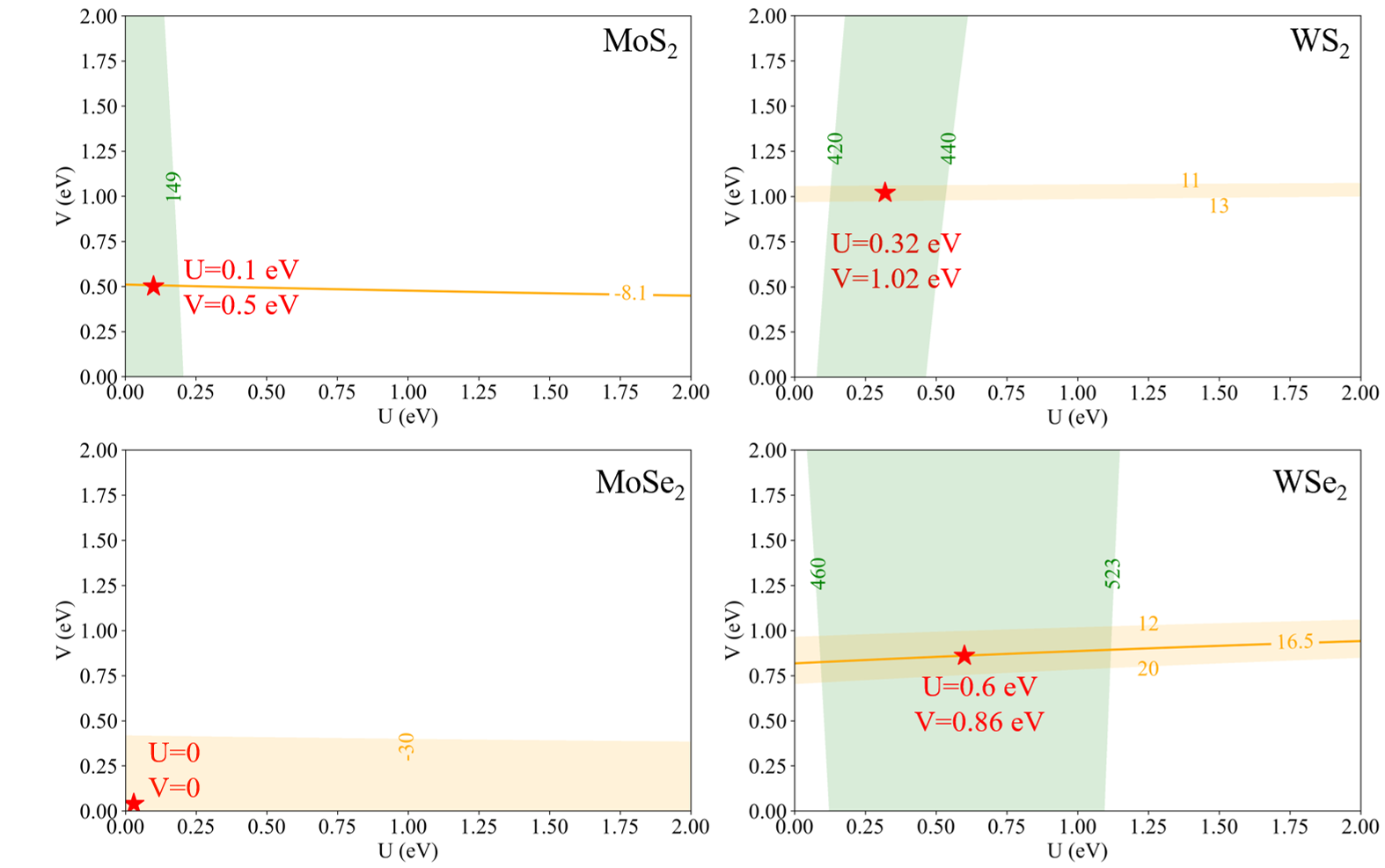}
\caption{Optimization of the Hubbard parameters $U$ and $V$ for WS$_2$, MoSe$_2$, and WSe$_2$. For MoS$_2$, we use previously optimized values~\cite{Rozhansky2026}. The shaded orange and green regions show the parameter spaces consistent with the experimental ranges of the CB ($\Delta_C$) and VB ($\Delta_V$) SO splittings, respectively. For MoSe$_2$ and WSe$_2$, the central CB benchmark is obtained from transport after removing the finite-density many-body enhancement; for WS$_2$, where a comparable transport benchmark is unavailable, it is taken from the optical value $\Delta_C=12$ meV. The VB benchmarks are taken from ARPES. The red stars mark the selected $U,V$ pairs corresponding to the central CB and VB benchmark values.}
   \label{fig:UV}
\end{figure*}

Overall, larger orbital angular momentum and stronger atomic SOC constants lead to enhanced spin splittings, while the relative orbital weights determine how these atomic SOC interactions project onto specific band edges. This explains why W-based TMDs (with heavier atoms and stronger SOC) exhibit much larger VB spin splitting and why the CB splitting remains small but material dependent, governed by $d-p$ hybridization effects, as the net splitting results from competing first- and second-order contributions of comparable magnitude. 

This also makes it clear why methods such as $G_0W_0$ calculations cannot generally resolve the discrepancy with experimental SO splitting, as the self-energy correction $\Sigma^{G_0W_0}-V_{\rm xc}$ is evaluated using the DFT eigenfunctions, and quasiparticle shifts are obtained as first-order corrections to the DFT energies, while metal--chalcogen orbital weights remain those of the underlying DFT calculation. At the same time, the CB and VB involve different orbital characters and SO mechanisms, so their splittings provide complementary constraints on the Hubbard parameters rather than a fit to a single observable. This motivates their use to determine $U$ and $V$.

\begin{table}[ht]
\centering
\small
\setlength{\tabcolsep}{2pt}
\caption{Experimentally established spin-orbit coupling parameters for CB and VB edges. SdHO identifies the extracted values from the analysis of Shubnikov--de Haas oscillations. (*) superscript marks recalculated values obtained upon taking into account exchange renormalisation of SO splitting at finite carrier densities, as described in this section and Ref.~\cite{Rozhansky2026}. Together with ARPES data, these are the benchmarks used to choose Hubbard parameters.}
\label{tab:exp}
\begin{tabular}{|c|c|c|}
\hline
 & $\Delta_{C}$ (meV): SdHO & $\Delta_{V}$ (meV): ARPES \\
\hline
MoS$_2$
& \begin{tabular}{c}
    $-15$\cite{PhysRevLett.121.247701} \\
    $\rightarrow\mathbf{-8.1}^{*}$\cite{Rozhansky2026}
  \end{tabular}
& \begin{tabular}{c}
    $145\pm4$\cite{PhysRevLett.114.046802}, $140\pm4$\cite{Lee2021}, \\
    $144$\cite{Ehlen2018}
  \end{tabular} \\
\hline
WS$_2$
& $12\pm1$\cite{PhysRevB.107.245407} (PL)
& \begin{tabular}{c}
    $431$\cite{PhysRevB.92.245442}, $420$\cite{Tanabe2016,Ulstrup2016}, \\
    $440$\cite{PhysRevB.97.155421}
  \end{tabular} \\
\hline
MoSe$_2$
& \begin{tabular}{c}
    $-37$\cite{PhysRevB.97.201407} \\
    $\rightarrow\mathbf{-19.5}^{*}$
  \end{tabular}
& $180$\cite{Zhang2013,Mo2016} \\
\hline
WSe$_2$
& \begin{tabular}{c}
    $29$\cite{Wang2016} \\
    $\rightarrow\mathbf{16.5}^{*}$
  \end{tabular}
& \begin{tabular}{c}
    $460$\cite{Zhao2019}, $469\pm8$\cite{PhysRevB.101.165103}, \\
    $475\pm5$\cite{Zhang2016}, $513\pm10$\cite{Le2015}
  \end{tabular} \\
\hline
\end{tabular}
\end{table}

\begin{table*}[ht]
\centering
\setlength{\tabcolsep}{2 pt}
\caption{Band structure parameters (left-hand side) recalculated with the optimally chosen Hubbard $U$ and $V$ (see Fig. 3), in comparison with the available experimentally established values (right-hand side). For comparison we also show the values DFT computed for $U=V=0$.}
\begin{tabular}{|c|c||c|c||c|c||c|c||c|c|}
\hline
  & \multirow{2}{*}{ $[U, V]$ ($eV$)} & 
  \multicolumn{2}{c||}{ $\Delta_{C}$ ($meV$)}
  & 
  \multicolumn{2}{c||}{ $\Delta_{V}$ ($meV$)}
  & \multicolumn{2}{c||}{ $E_g$ ($eV$)} & 
  \multicolumn{2}{c|}{$\Delta_{FE}$($mV$)}
   \\
\cline{3-10}
  &   &  theory &  exp. &  theory &  exp. & theory &  exp. & theory &  exp. \\
\hline

\multirow{2}{*}{MoS$_2$}  & [0,0] & -2.7 & \multirow{2}{*}{-8.1 \cite{
Rozhansky2026}} & 145.8 & \multirow{2}{*}{136-149 \cite{PhysRevLett.114.046802, Lee2021, Ehlen2018}} 
& 1.71 & \multirow{2}{*}{2.106-2.16\cite{PhysRevB.110.235413, Vaquero2020, PhysRevLett.123.136801}}& 70 & \multirow{2}{*}{58$\pm$1.5\cite{PhysRevX.12.041005}, 47\cite{Wang2022}}\\

& [0.1,0.5] &  -8 &  &  148 & & 1.83 &  & 62 & \\
\hline
\multirow{2}{*}{WS$_2$}  & [0,0] &  36.6 &  \multirow{2}{*}{12\cite{PhysRevB.107.245407}} & 417.7 & \multirow{2}{*}{420-440 
\cite{PhysRevB.92.245442, Tanabe2016,Ulstrup2016, PhysRevB.97.155421}} 
& 1.68 & \multirow{2}{*}{2.3-2.45 \cite{PhysRevLett.113.076802, Hsu2019}} & 67 & \multirow{2}{*}{56$\pm$3\cite{Wang2022}}\\
& [0.32,1.02]    & 11.1 &  & 429.7 & &  1.89& & 56 & \\
\hline

MoSe$_2$ & [0,0] &  -21.9 & -19.5$^*$\cite{PhysRevB.97.201407} & 183.7 & 180 \cite{Zhang2013,Mo2016} & 1.45 & 1.648\cite{Viner2021} & 58 & 57$\pm$5\cite{Wang2022}\\

\hline

\multirow{2}{*}{WSe$_2$}  & [0,0] &  44.8 & \multirow{2}{*}{16.5$^*$ \cite{Wang2016}} &  454.3 & \multirow{2}{*}{460-513 \cite{Zhao2019, PhysRevB.101.165103, Zhang2016, Le2015}} 
& 1.39 & \multirow{2}{*}{1.74-1.894\cite{Viner2021, Kapuciski2021}} & 56 & \multirow{2}{*}{56$\pm$5\cite{Wang2022}}\\
&[0.6,0.86] & 16.1 &  & 490.3 & & 1.53 &  &  48 &\\
\hline
\end{tabular}
\label{tab:SOS}
\end{table*}

\subsection{Optimizing $U$ and $V$ from experimental spin-orbit measurements}
In order to optimise the choice of $U$ and $V$ parameters, we require experimentally measured values of the single-particle spin--orbit splitting at the bottom of the CB and the top of the VB. We have therefore surveyed the available experimental literature and selected those measurements that provide the most reliable estimates of these properties. For the top of the VB we use the values directly provided by ARPES measurements. For the CB, we use the SO splitting obtained from the analysis of the Shubnikov--de Haas oscillations (SdHO), which tell us the critical electron density $n^*$, above which the upper spin-split CB gets filled. At this finite electron density, the band edge energy has a contribution from many-body electron-electron exchange interaction, thereby enhancing the extracted CB SO splitting relative to the single-particle value. We subtract this enhancement following the approach developed in Refs.~\cite{PhysRevB.110.L161404,Rozhansky2026}. In particular, we calculate the electron self-energy exchange term with account for the screened Keldysh interaction using the interaction-corrected long-wavelength polarisation. Further details can be found in the SI. The resulting single-particle CB SO splittings are listed in Table ~\ref{tab:exp} in bold for MoS$_2$, MoSe$_2$, WSe$_2$ (together with the ones naively estimated from $n^*$)  and used as the input for $U$ and $V$ optimisation. For WS$_2$, where SdHO data are unavailable, we use SO splitting extracted from the analysis of dark-bright exciton splitting.

To match the selected values of VB and CB splitting with our DFT data, we extract their values on a finite grid of $U$ and $V$, before interpolating the calculated values to produce a smooth representation of their dependence on the Hubbard parameters, as detailed in the SI. Fig.~\ref{fig:UV} shows regions where the calculated splittings agree with the experimental ranges defined above. The widths of these regions reflect the spread and uncertainty of the experimental data, and we therefore choose the optimal values as those sets of Hubbard parameters which lie within the central overlapping regions of CB and VB experimental ranges.

Before continuing, we further compare the calculated values of the monolayer band gap, $E_g$, at the identified $U, V$ values to experiment. To compare, we use experimental values extracted from excitonic Rydberg-series measurements, for which fitting the series towards its $n\rightarrow\infty$ limit provides an estimate of the quasiparticle band gap without requiring an independent determination of the exciton binding energy. As shown in Table~\ref{tab:SOS}, inclusion of non-zero $U$ and $V$ also systematically shifts calculated band gaps towards the experimental values, in each case where a Hubbard correction is applied, despite this not being used in the choice of $U$ and $V$ values.

\subsection{Band-edge energetics and effective masses}
\begin{figure*}[ht]
\centering    
\includegraphics[width=2\columnwidth]{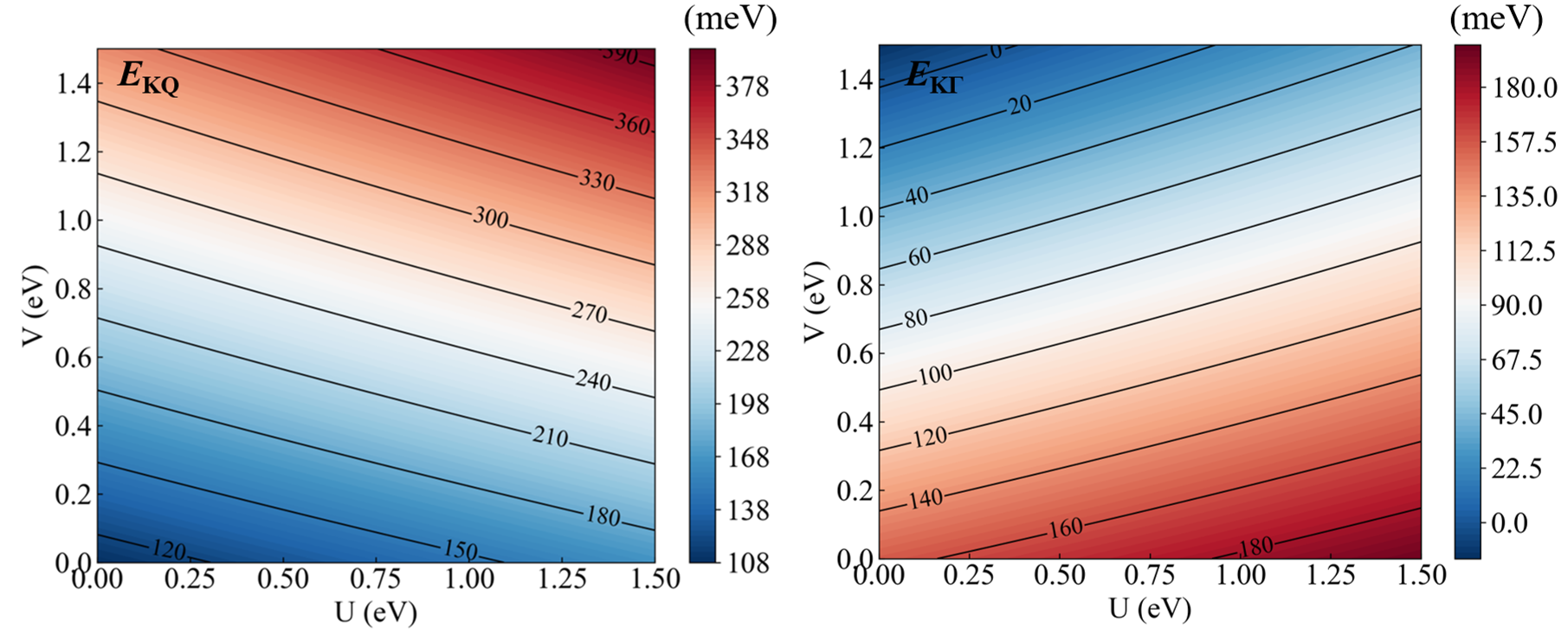}
\caption{Contour maps of band edge offsets of monolayer MoS$_2$ as functions of the Hubbard interaction strengths $U$ and $V$.}
   \label{fig:energymono}
\end{figure*}

We additionally compare several additional electronic structure properties, including band-edge energy offsets and carrier effective masses, for our optimized Hubbard parameters.  As shown in Fig.~\ref{fig:energymono}, the energy difference $E_{KQ}$ increases with both interaction parameters, with a substantially stronger sensitivity to $V$. The optimal value in MoS$_2$ is in again a better match to experiment, as it produces a closer agreement with our recent calculations of layer doping in 4-layer MoS$_2$, where a larger offset was required to match experiment \cite{Masseroni2025}. In contrast, $E_{K\Gamma}$ decreases with increasing $V$ and shows only a weak dependence on $U$. We have also calculated the dependence of electron masses at the $K$ and $Q$ valleys, which generally vary with both the $U$ and $V$ parameters, whereas the hole masses are predominantly controlled by $V$. In all cases, the overall variation in mass remains relatively small, particularly over the range of experimentally measured TMD masses, suggesting that the optimized Hubbard parameters mainly affect band-edge energetics and hybridization rather than carrier transport properties.

\subsection{Ferroelectric polarization in parallel-stacked TMD bilayers}
For TMD bilayers, the stacking orientation of each monolayer determines the overall symmetry of the system, giving two distinct configurations: centrosymmetric hexagonal (2H) stacking, with an inversion centre located between the layers, and rhombohedral (3R) stacking, which lacks this inversion centre and permits interlayer charge transfer \cite{Strachan2021, PhysRevB.95.115429, ViznerStern2021, Enaldiev2021}. This broken symmetry produces a spontaneous out-of-plane polarisation and associated electrostatic potential drop across 3R domains\cite{Li2017, Yang2023, ferreira2022scaleability, Heidkamp2026, Ferreira2021}, underpinning a novel form of ferroelectricity, where different polarisations can be switched by sliding of the layers.

\begin{figure}[hb]
\centering    
\includegraphics[width=1\columnwidth]{ 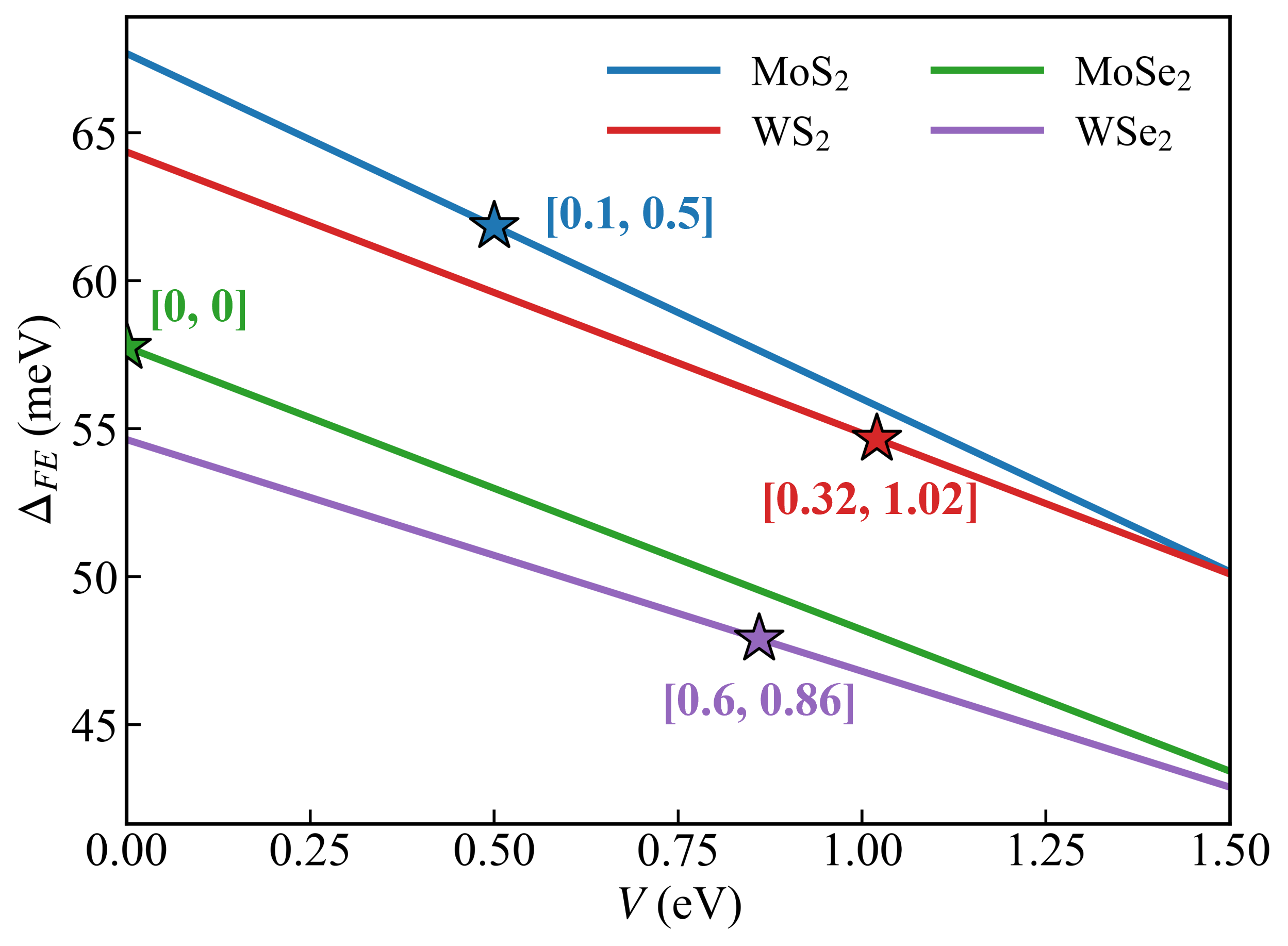}
\caption{Dependence of the interlayer potential drop on the intersite Coulomb interaction $V$, for optimal on-site $U$, for the four TMDs studied in this work.}
   \label{fig:drop}
\end{figure}

Importantly, this interlayer potential drop is strongly related to hybridisation between chalcogen orbitals on adjacent layers, and may therefore be susceptible to the changes in orbital composition outlined above. Moreover, $\Delta_{\rm FE}$ provides a useful independent benchmark for the Hubbard correction, since the built-in interlayer potential can be determined experimentally through electrical and optical measurements independently of the SO splittings used to optimise $U$ and $V$ \cite{Wang2022,PhysRevX.12.041005}. We therefore apply the identified optimal Hubbard parameters to inversion-symmetry-breaking, parallel-stacked bilayers, which host an out-of-plane polarisation. We find that, while $\Delta_{\rm FE}$ has only a weak dependence on $U$, it decreases monotonically with increasing $V$, as shown in Fig.~\ref{fig:drop}. Consequently, the optimised Hubbard parameters systematically reduce the larger values obtained with standard DFT towards experiment, giving substantially improved agreement with the measured potential drops reported in Table~\ref{tab:SOS}.

\subsection{Interlayer hybridization in TMD bilayers}
As a final benchmark, we also examine how the band edge energies evolves when they are stacked into bilayers. In bilayers, interlayer coupling also modifies splitting and offsets, as compared to a monolayer. This can be characterized by an effective interlayer hybridization parameter $t$, which governs the mixing of layer-resolved states and the resulting band splittings. To quantify the influence of stacking geometry and electronic interactions on interlayer coupling, we employ a minimal two-level model for the bilayer VB states. In the basis of top- and bottom-layer states, Hamiltonians for both spins can be written as:

\begin{equation}
H=
\begin{pmatrix}
\epsilon_k+\tau S^z \dfrac{\Delta_{\rm SO}}{2}
+\dfrac{\Delta_{\rm FE}}{2} & t \\[6pt]
t &
\epsilon_k+\tau\eta S^z \dfrac{\Delta_{\rm SO}}{2}
-\dfrac{\Delta_{\rm FE}}{2}
\end{pmatrix},
\end{equation}
where $\epsilon_k=\hbar^2k^2/2m$. Here, $t$ denotes the interlayer hybridization strength, $\Delta_{\rm SO}$ is the monolayer SO splitting, and $\Delta_{FE}$ represents the electrostatic potential difference between the two layers. Index $\tau$ labels $\pm K$ valley, while $\eta$ corresponds to stacking orientation, which is $1$ for 3R and $-1$ for 2H bilayers. This gives the set of eigenvalues, which we can directly fit to our DFT calculations.

\begin{align}
&E_{\pm}^{\rm 2H}
=
\frac{\hbar^2k^2}{2m}
\pm
\sqrt{
\frac{\Delta_{\rm SO}^2}{4}
+t^2
},\nonumber\\
&E_{\pm}^{\rm 3R}
=
\frac{\hbar^2k^2}{2m}
+
\frac{S^z\Delta_{\rm SO}}{2}
\pm
\sqrt{
\frac{\Delta_{\rm FE}^2}{4}
+t^2
}.
\end{align}

\begin{figure*}[ht]
\centering    
\includegraphics[width=2\columnwidth]{ 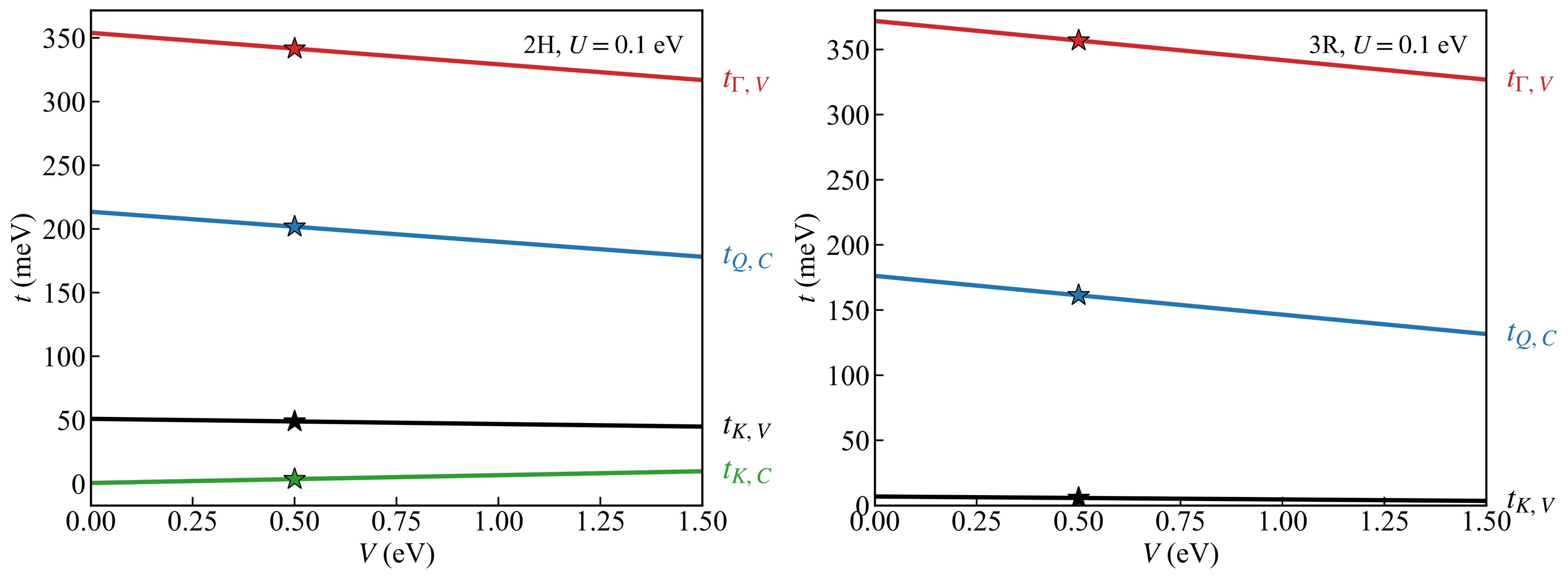}
\caption{Hybridization strength of different band edges in 2H- and 3R-MoS$_2$ bilayers\cite{PhysRevB.95.115429}, for optimal $U=0.1$ eV. Material parameters at optimal Hubbard values are indicated by star symbols.}
   \label{fig:hybridization}
\end{figure*}

Fig.~\ref{fig:hybridization} shows the variation in hybridisation parameter for all band edges.
For $\Gamma$ and $Q$ valleys, which have a more substantial out-of-plane orbital character, the hybridization decreases monotonically with increasing $V$, while $K$ point $t_{K,V}$ and $t_{K,C}$ exhibit only a weak dependence. We have provided the full set of eigenvalues for all band edges with and without a Hubbard correction in the SI. 

For further reference, we provide in the SI full parameterisation of every quantity which we have extracted from Hubbard-dependent DFT, in this work. This also includes calculations of various additional parameterisations, and associated plots, for other physically important quantities including in-plane and out-of-plane lattice constants, and monolayer polarizability. Most notably, we find that while our Hubbard optimization procedure improves the electronic properties which we discussed throughout the text, it has minimal impact on these structural parameters, which are naturally less sensitive to finer details of the orbital composition of individual states.

\section{Discussion}
In this work, we have shown that the extended DFT+$U$+$V$ method provides a practical route to improve the description of band-edge hybridisation and, consequently, spin--orbit splitting in TMDs. In contrast to its more conventional application to strongly correlated materials, where Hubbard corrections are often introduced to correct substantial localisation errors, here we find that the $U$ and $V$ terms can allow for comparatively fine control over the orbital composition of otherwise well-described semiconductor band edges, as relatively small changes in $M$-$d$--$X$-$p$ hybridisation can nevertheless have an outsized influence on quantities such as the CB spin--orbit splitting. 

We also find that these parameters improve the description of various other material properties, that were not used during calibration, including band-edge offsets and the stacking-induced electrostatic potential drop in rhombohedral bilayers. This transfer between different properties and multilayer configurations suggests that the experimentally calibrated corrections identified here partially correct a common underlying error in the standard-DFT description of inter-site hybridisation between $M$-$d$ and $X$-$p$ states. Hubbard corrections have the further advantage of being extremely computationally efficient: the approach employed in this work can be immediately applied to produced quantitatively accurate predictions for much larger systems, such as moir\'e structures, complicated multilayer stacks, or cells containing point defects, where systematic hybrid-functional or full $GW$ calculations are generally prohibitively expensive.

A natural future application of this approach is in the refinement of computational two-dimensional materials databases, as it is likely that many properties of other 2D materials are sensitive to the orbital composition. Existing databases typically only report properties calculated using conventional semilocal functionals. Supplementing this data, with calculations on a limited grid of $U,V$ values, along with interpolation of selected electronic descriptors, could provide material- and environment-dependent corrections while retaining the efficiency required for high-throughput screening. This type of minimal interpolation also supports future parameterisation, based on  experimental measurements or more expensive reference calculations. 

\section*{Methods}
\subsection*{DFT and DFT+$U$+$V$ calculations}
First-principles calculations were performed using the Quantum ESPRESSO (QE) package. The exchange--correlation energy was treated within the generalized gradient approximation (GGA), using the Perdew--Burke--Ernzerhof (PBE) functional, together with fully relativistic ultrasoft pseudopotentials for all atomic species. The Kohn--Sham wavefunctions and charge density were expanded in a plane-wave basis with kinetic-energy cutoffs of 80 Ry and 800 Ry, respectively. Brillouin-zone integrations employed a uniform $31\times31\times1$ $k$-point mesh for structural optimization and self-consistent calculations, with electronic occupations treated using Fermi--Dirac smearing of width 0.001 Ry. A vacuum spacing of 15~$\text{\AA}$ was introduced along the out-of-plane direction to suppress interactions between periodically repeated monolayers. Van der Waals interactions were included using the vdW-DF2-c09 functional to provide an accurate description of interlayer binding in multilayer structures. Atomic positions were relaxed using the Broyden--Fletcher--Goldfarb--Shanno (BFGS) quasi-Newton algorithm until the residual forces on all atoms were below $0.01$~eV/$\text{\AA}$.

Spin--orbit coupling (SOC) was included self-consistently within the non-collinear formalism using the fully relativistic pseudopotentials, and all electronic-structure and band-dispersion calculations reported in this work include SOC unless stated otherwise. Extended DFT+$U$+$V$ calculations were performed using the ortho-atomic projector formalism implemented in QE, with the same plane-wave cutoffs, $k$-point sampling, smearing, and structural parameters as the corresponding conventional DFT calculations. For each $(U,V)$ combination, the electronic ground state was determined fully self-consistently. The on-site Hubbard interaction $U$ applied to the transition-metal $d$ orbitals was systematically varied from $0$ to $2$~eV in increments of $0.5$~eV, while the inter-site interaction $V$ between metal $d$ and symmetry-inequivalent neighbouring chalcogen $p$ orbitals was varied from $0$ to $2$~eV in increments of $0.1$~eV. This two-dimensional parameter sweep was used to quantify the dependence of the band-edge energies, orbital hybridisation, and resulting spin--orbit splittings on the local and inter-site Hubbard corrections.

\section*{Data availability}
The data supporting the findings of this study are available from the corresponding author upon reasonable request.

\section*{Acknowledgements}
We thank Alexey Chernikov for fruitful discussion. X.L. acknowledges financial support from the University of Manchester’s Dean’s Doctoral Scholarship. 
S.A. acknowledges the support of the Deanship of Scientfic Research at Northern Border University, Arar, KSA for funding through the project number NBU-SA-FIR-2025. J.G.M. is supported by the University of Manchester Dame Kathleen Ollerenshaw Fellowship. The authors acknowledge the use of resources provided by the Isambard 3 Tier-2 HPC Facility, hosted by the University of Bristol and operated by the GW4 Alliance (\url{https://gw4.ac.uk}) and funded by UK Research and Innovation; and the Engineering and Physical Sciences Research Council [EP/X039137/1].
We acknowledge support from 2DFERROPLEX project funded by the European Union under grant agreement 101223116.

\section*{Author contributions}
V.F. conceived the idea of the project and designed its methodology. X.L. and J.G.M. performed DFT modelling and determined input parameters for the DFT+$U$+$V$ method and the conduction and valence band spin-orbit splitting were fitted by S.A. and I.R. All authors wrote the manuscript.

\section*{Competing Interests}
The authors declare no competing interests.

\bibliography{references}

\end{document}